\documentclass{iauc}
\usepackage{graphicx}
\pubyear{2005}
\volume{199}
\pagerange{1--}
\jname{Probing Galaxies through Quasar Absorption Lines}
\editors{P. R. Williams, C. Shu, and B. M\'{e}nard, eds.}

\title[The mass-metallicity relation for high-$z$ DLA galaxies] 
{The mass-metallicity relation for high-redshift damped Ly$\alpha$ galaxies}

\author[C. Ledoux et al.] 
{C\'edric Ledoux$^1$,
Patrick Petitjean$^2$, Palle M\o ller$^3$, \break Johan Fynbo$^4$,
\and R. Srianand$^5$}

\affiliation{$^1$European Southern Observatory (ESO), Alonso de C\'ordova 3107,
Vitacura, Santiago, Chile\break
email: cledoux@eso.org\\[\affilskip]
$^2$Institut d'Astrophysique de Paris -- CNRS, 98bis Boulevard Arago,
75014, Paris, France\\[\affilskip]
$^3$ESO, Karl-Schwarzschild-Stra\ss e 2, 85748, Garching bei
M\"unchen, Germany\\[\affilskip]
$^4$Astronomical Observatory, Juliane Maries Vej 30, 2100 Copenhagen $\O$, Denmark\\[\affilskip]
$^5$IUCAA, Post Bag 4, Ganesh Khind, Pune 411 007, India}

\begin{document}

\maketitle

\begin{abstract}
We used our database of ESO VLT-UVES spectra of quasars to build up a sample of
67 Damped Lyman-$\alpha$ (DLA) systems with redshifts $1.7<z_{\rm abs}<3.7$.
For each system, we measured average metallicities relative to Solar, [X/H]
(with either ${\rm X}={\rm Zn}$, S or Si), and the velocity widths of
low-ionization line profiles, $W_1$. We find that there is a tight correlation
between the two quantities, detected at the $5\sigma$ significance level. The
existence of such a correlation, over more than two orders of magnitude
spread in metallicity, is likely to be the consequence of an underlying
mass-metallicity relation for the galaxies responsible for DLA absorption
lines. The best-fit linear relation
is $[{\rm X}/{\rm H}]=1.35(\pm 0.11)\log W_1 -3.69(\pm 0.18)$
with $W_1$ expressed in km s$^{-1}$. While the slope of
this velocity-metallicity relation is the same within uncertainties between
the higher and the lower redshift bins of our sample, there is a hint of an
increase of the intercept point of the relation with decreasing redshift. This
suggests that galaxy halos of a given mass tend to become more metal-rich
with time. Moreover, the slope of this relation is consistent with that of the
luminosity-metallicity relation for local galaxies. The DLA systems having
the lowest metallicities among the DLA population would therefore, on
average, correspond to the galaxies having the lowest masses. In turn, these
galaxies should have the lowest luminosities among the DLA galaxy
population. This may explain the recent result that the few DLA systems
with detected Ly$\alpha$ emission have higher than average metallicities.
\end{abstract}


\section{Observations}

Most of the DLA systems in our sample (with total neutral hydrogen
column densities $\log N($H\,{\sc i}$)\ge 20$) were observed at the ESO
Very Large Telescope (VLT) with UVES between 2000 and 2003 to search for H$_2$
molecules at $z_{\rm abs}>1.8$ (Petitjean et al. 2000; Ledoux et al. 2003). The
absorption line analysis was homogeneously performed using standard
Voigt-profile fitting techniques. Average DLA metallicities relative to Solar,
[X/H$]\equiv\log [N($X$)/N($H$)]-\log [N($X$)/N($H$)]_\odot$, were
calculated as the sum of the column densities measured in individual components
of the absorption profiles, with X$=$Zn as the reference element when
Zn\,{\sc ii} is detected, or else either S or Si. One notable
characteristics of this large dataset is that it samples well both the low and
the high ends of the DLA metallicity distribution, from [X/H$]\approx -2.6$
up to about half of Solar.

For each DLA system, we also measured the velocity widths of the
metal absorption line profiles. Low-ionization transition lines that are
not strongly saturated were selected (see
Prochaska \& Wolfe 1997; Wolfe \& Prochaska 1998). For high-ionization lines,
the velocity width could be dominated by extended galaxy halos and galactic
winds. We calculated the velocity width of the selected profiles as the
second moment of the distribution of apparent optical depth along
these profiles.

\begin{figure}
\center
\includegraphics[bb=45 53 569 628,clip,width=9.cm,angle=-90.]{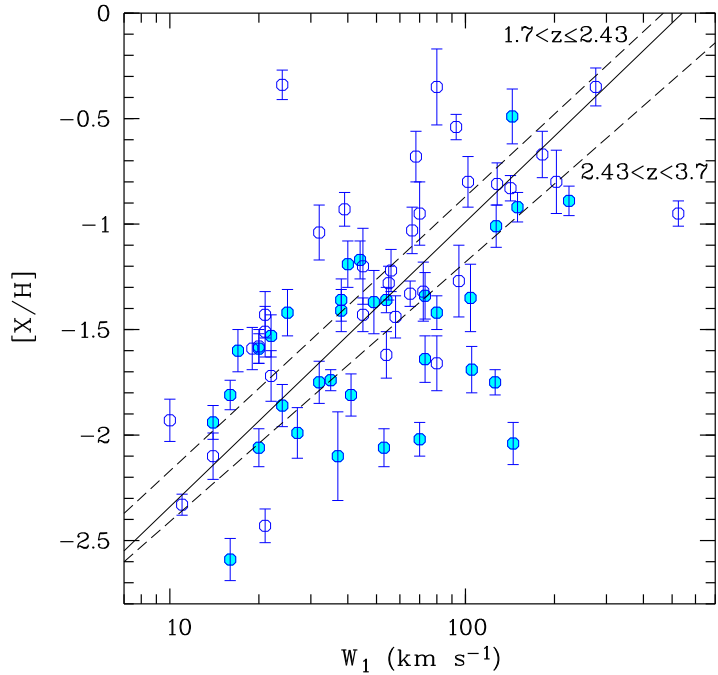}
\caption{Average metallicity of each DLA system in our
sample, [X/H], vs. $W_1$ on a logarithmic scale (empty
circles: $1.7<z_{\rm abs}\le 2.43$; shaded circles: $2.43<z_{\rm abs}<3.7$). A
correlation between the two quantities is detected at
the $5\sigma$ significance level. The linear least-squares bisector fit to
the total sample (resp. the $z_{\rm abs}\le 2.43$ or $z_{\rm abs}>2.43$
sub-samples) is shown by a solid line (resp. long-dashed lines).}
\label{fig1}
\end{figure}

\section{DLA velocity-metallicity correlation}

The correlation between DLA low-ionization metal line width and
average metallicity, as observed in Fig.~1, over more than two orders
of magnitude spread in metallicity, can be understood as the consequence of
an underlying mass-metallicity relation for the galaxies responsible for
DLA absorption lines. We fitted the data using the linear least-squares
bisector method. For DLA systems with redshifts larger (resp. smaller) than the
median redshift of our sample, we
find $[{\rm X}/{\rm H}]=1.23(\pm 0.14)\log W_1 -3.64(\pm 0.24)$
at $2.43<z_{\rm abs}<3.7$ (resp.
$[{\rm X}/{\rm H}]=1.30(\pm 0.16)\log W_1 -3.47(\pm 0.26)$ at
$1.7<z_{\rm abs}\le 2.43$).

While the slope of the DLA velocity-metallicity relation is the same
within uncertainties between the higher and the lower redshift bins of our
sample, there is a hint of an increase of the intercept point of the
relation with decreasing redshift. This is linked to an increase of the
mean DLA metallicity with decreasing redshift, with [X/H$]=-1.58$
(resp. [X/H$]=-1.22$) in the higher (resp. lower) redshift bin of our sample
(see also Prochaska et al. 2003). {\it This suggests that galaxy halos of
a given mass (resp. a given metallicity) tend to become more metal-rich (resp.
less massive) with time}. This is consistent with the recent result by
Savaglio et al. (2004) for $0.4<z<1$ galaxies selected from the Gemini Deep
Deep Survey and the Canada-France Redshift Survey.

The above results are also in agreement with those of Nestor et al. (2003)
who found larger metallicity in Sloan Digital Sky
Survey (SDSS) $0.9<z_{\rm abs}<2.2$ DLA composites with
larger Mg\,{\sc ii}$\lambda$2796 equivalent widths. In addition, these
authors showed that, within the large equivalent width regime, the
metallicity is larger at lower redshift.

\section{Implications and prospects}

Haehnelt et al. (1998) performed similar line velocity width measurements
on low-ionization lines from DLA systems in simulations and compared them
directly to the virial velocities of the underlying dark matter
halos ($v_{\rm vir}=(GM/r)^{1/2}$ in a sphere overdense by a factor of
200 compared to the mean cosmic density). They found $W_1=0.4v_{\rm vir}$.
According to Haehnelt et al. (2000), the luminosity function of $z\sim 3$
galaxies can be reproduced if a simple linear scaling of the luminosity
with the mass of the dark matter halos is assumed, i.e.,
$m_B=-7.5\log (v_{\rm vir}/200\ {\rm km\ s}^{-1})+m_B^0$, where $m_B$ is the
galaxy apparent $B$-band magnitude and $m_B^0\simeq m_R^0=26.6$. Using
our linear best-fit to the velocity-metallicity relation for
$1.7<z_{\rm abs}<3.7$ DLA systems ($z_{\rm med}=2.43$), we get:
\begin{equation}\label{eq1}
[{\rm X}/{\rm H}]=-0.18(\pm 0.02)M_B -4.70(\pm 0.18)-0.18K_B,
\end{equation}
where $K_B$ is the $K$-correction in the $B$-band, which should be positive.
The slope of this DLA luminosity-metallicity relation is consistent with that
derived by Tremonti et al. (2004) for the luminosity-metallicity relation for
$z\sim 0.1$ galaxies selected from the SDSS,
$[{\rm O}/{\rm H}]=-0.185(\pm 0.001)M_B-3.452(\pm 0.018)$. However, the
zero points of the two relations are different {\it implying that galaxies of
a given luminosity (resp. a given metallicity) are becoming more metal-rich
(resp. fainter) with time}.

Eq.~\ref{eq1} implies that the more than two orders of magnitude spread in DLA
metallicity reflects a more than ten magnitudes spread in DLA
galaxy luminosity. In other words, a low metallicity should on average imply
a small stellar mass and thus a low luminosity. Even though low-mass galaxies,
i.e., gas-rich dwarf galaxies, can undergo periods of intense star
formation activity, they show, on average, lower star-formation rates than more
massive galaxies (Brinchmann et al. 2004; see also Okoshi et al. 2004).
The existence of a DLA mass-metallicity relation may thus explain the fact that
the few DLA systems with detected Ly$\alpha$ emission have higher than
average metallicities (M\o ller et al. 2004; Christensen et al., in prep.).
This should be confirmed by additional deep imaging of the fields of QSOs with
selected DLA absorbers.

\end{document}